\documentclass[a4paper,fleqn,usenatbib]{mnras}

\usepackage{color}
\usepackage{graphicx}
\usepackage{amsmath}
\usepackage{amssymb}

\title[F-GAMMA:  Variability Doppler factors]{F-GAMMA:  Variability Doppler factors of blazars from multiwavelength monitoring}
\author[Liodakis et al.]
{I. Liodakis$^{1,2}$\thanks{liodakis@physics.uoc.gr}, N. Marchili$^{4}$, E. Angelakis$^{3}$, 
L. Fuhrmann$^{3,6}$, I. Nestoras$^{3}$, I. Myserlis$^{3}$,\newauthor V. Karamanavis$^{3}$, T. P. Krichbaum$^{3}$,  A. Sievers$^{5}$, H. Ungerechts$^{5}$, J. A. Zensus$^{3}$\\
$^{1}$Department of Physics and ITCP\thanks{Institute for Theoretical and Computational Physics, formerly Institute for Plasma Physics}, University of Crete, 71003, Heraklion, Greece\\
$^{2}$Foundation for Research and Technology - Hellas, IESL, Voutes, 7110 Heraklion, Greece\\
$^{3}$Max-Planck-Institut f\"ur Radioastronomie, Auf dem Ḧ\"ugel 69, 53121 Bonn, Germany\\
$^{4}$IAPS-INAF, Via Fosso del Cavaliere 100, 00133, Roma, Italy\\
$^{5}$Instituto de Radio Astronom\'{i}a Milim\'{e}trica, Avenida Divina Pastora 7, Local 20, 18012 Granada, Spain\\
$^{6}$ZESS - Center for Sensorsystems, University of Siegen, Paul-Bonatz-Str. 9-11, 57076 Siegen, Germany\\
}

\begin{document}
\maketitle
\label{firstpage}
\begin{abstract}
Recent population studies have shown that the variability Doppler factors can adequately describe blazars as a population. We use the flux density variations found within the extensive radio multi-wavelength datasets of the F-GAMMA program, a total of 10 frequencies from 2.64 up to 142.33 GHz, in order to estimate the variability Doppler factors for 58  $\gamma$-ray bright sources, for 20 of which no variability Doppler factor has been estimated before. We employ specifically designed algorithms in order to obtain a model for each flare at each frequency. We then identify each event and track its evolution through all the available frequencies for each source. This approach allows us to distinguish significant events producing flares from stochastic variability in blazar jets. It also allows us to effectively constrain the variability brightness temperature and hence the variability Doppler factor as well as provide error estimates. Our method can produce the most accurate (16\% error on average) estimates in the literature to date.
\end{abstract}


\section{Introduction}\label{introdu}

Blazars, the sub-class of active galactic nuclei (AGN) with their jet axis pointing towards us, includes the Flat Spectrum Radio Quasars (FSRQs) and BL Lac objects that dominate the $\gamma$-ray extragalactic sky. Blazars are characterized by extremely broad-band emission (from long cm radio wavelengths to TeV energies), intense variability at all wavelengths, relativistic boosting of the emitted luminosity and often significantly apparent superluminal motion. Most of these exotic phenomena are attributed to the combination of the relativistic speeds and the alignment of the jet to our line of sight \citep{Blandford1979}, which obscure our view of their intrinsic properties. The observed properties of blazar jets are modulated by the Doppler factor defined as $\delta=[\Gamma(1-\beta\cos\theta)]^{-1}$, where $\Gamma=(\sqrt{1-\beta^2})^{-1}$ is the Lorentz factor, $\beta$ the velocity of the jet in units of speed of light, and $\theta$ the jet viewing angle.

Being one of the most important parameters in the blazar paradigm many methods have been proposed for estimating $\delta$. Such methods are the equipartition Doppler factors  \citep{Readhead1994,Guijosa1996}, the variability Doppler factors   \citep{Valtaoja1999,Lahteenmaki1999-III}, the single component causality Doppler factors  \citep{Jorstad2005,Jorstad2006}, as well as the inverse Compton Doppler factors \citep{Ghisellini1993}, and the $\gamma$-ray opacity Doppler factors \citep{Mattox1993,Dondi1995}. The equipartition and variability Doppler factors are based on the assumption of equipartition  between the energy density of the magnetic field and the radiating particles \citep{Readhead1994}. The former uses the brightness temperature measured from VLBI observations while the latter the variability brightness temperature from flux density variations. The single component causality Doppler factor method uses the observed angular size and variability timescale to calculate the Doppler factor for each individual component. The Doppler factor of a source is then calculated as the weighted mean of the Doppler factors of all the components, with weights inversely proportional to the uncertainty in the apparent velocity of each component. The inverse Compton Doppler factors use the framework of the Synchrotron Self-Compton (SSC) model in order to estimate the expected X-ray flux density given the angular size and flux density of the core from VLBI observations. The Doppler factor is obtained by comparing the observed and the theoretically expected X-ray flux density. The $\gamma$-ray opacity Doppler factors use pair production absorption effects resulting from the interaction of $\gamma-$ and X-rays. Assuming that the emission region has a spherical geometry, that X-rays and $\gamma$-rays are co-spatial and that the region is transparent to $\gamma$-rays, a lower limit of the Doppler factor can be obtained by relating the variability timescale to the size of the emission region.

Each one of the above methods is using different assumptions, that might not hold. Thus a direct comparison of the results from different methods is unable to provide the answer to which method can best describe blazars. Recent population models \citep{Liodakis2015} have shown that the variability Doppler factor method \citep{Valtaoja1999,Lahteenmaki1999-II,Lahteenmaki1999-III,Hovatta2009} can adequately describe both the FSRQ and BL Lac populations \citep{Liodakis2015-II}, although application on a source-by-source basis has to be performed with caution. Moreover, an error analysis has shown that although it is the most accurate (~30\% on average error on each estimate), it suffers from systematics introduced due to the cadence of observations. Since the method involves fitting with exponentials the flux density radio light curves (in order to calculate the variability timescale, flare amplitude, and then the variability brightness temperature), flares faster than the cadence of observations will be unresolved, setting an upper limit to the fastest observed timescale and thus the Doppler factor. We can overcome such limitation in two ways. Either by using data from surveys with high cadence observations such as the Owens Valley Radio Observatory (OVRO)\footnote{http://www.astro.caltech.edu/ovroblazars/} blazar program \citep{Richards2011} or, as in our case, by modeling the flares.

In this work we use the extensive 8-year-long multi-wavelength radio light curves from the F-GAMMA program\footnote{http://www3.mpifr-bonn.mpg.de/div/vlbi/fgamma/fgamma.html} \citep{Fuhrmann2007,Angelakis2010,Angelakis2012,Fuhrmann2016}. The F-GAMMA program monitored a sample of powerful and variable sources detected by the Fermi gamma-ray space telescope\footnote{http://fermi.gsfc.nasa.gov/} \citep{Acero2015} at ten frequencies from 2.64 up to 142.33 GHz  with an approximately monthly cadence (sparse datasets at 228.9 and 345 GHz are also available). Our goals were to distinguish significant events occurring in blazar jets from stochastic variations, and effectively constrain the variability parameters  of each source in order to estimate their variability Doppler factor. The method we use to estimate the variability Doppler factors is described in detail in \cite{Angelakis2015}. For the purposes of the current work, an error estimation step has been added in our analysis pipeline.

The paper is organized as follows. In Section \ref{method} we give a short description of the methods used. In section \ref{Variability Doppler factors} we present our estimates for the variability Doppler factors, Lorentz factors, and viewing angles of the sources in our sample, in Section \ref{compar} a comparison with estimates from the literature which we use as a proxy to validate our estimates, and in Section \ref{summ} we summarize our results.

The cosmological parameters we adopt in this work are $H_0=71$ ${\rm km \, s^{-1} \, Mpc^{-1}}$, $\Omega_m=0.27$ and $\Omega_\Lambda=1-\Omega_m$ \citep{Komatsu2009}.

\section{Methods}\label{method}
\begin{figure}
\resizebox{\hsize}{!}{\includegraphics[scale=1,angle =0]{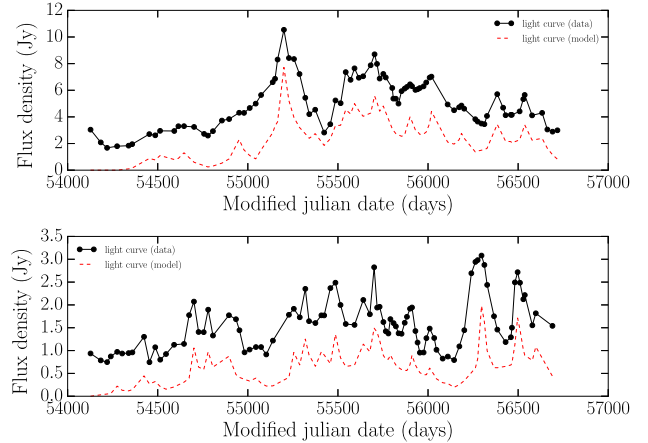}  }
 \caption{Observed (solid black) and simulated (dashed red) light curves for OJ287 (15~GHz, upper panel) and J0721+7120 (10.45~GHz, lower panel) after the flare modeling procedure has been completed.}
 \label{plt_sim_light}
 \end{figure}
\begin{figure}
\resizebox{\hsize}{!}{\includegraphics[scale=1,angle =0]{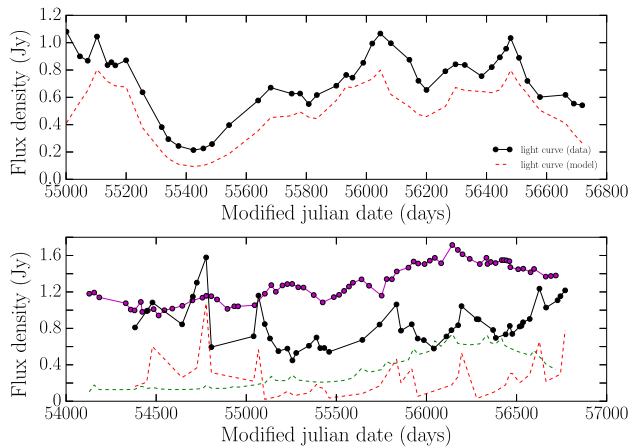}  }
 \caption{Observed (solid black) and simulated (dashed red) light curves for J0050-0929 (8.35~GHz, upper panel) and J0241-0815 (142.33 GHz solid black and dashed red, 10.45~GHz solid magenta and dashed green) after the flare modeling procedure has been completed.}
 \label{plt_sim_light2}
 \end{figure}

The calculation of the variability brightness temperatures and Doppler factors of our sources depends on the estimation of their variability characteristics, i.e. the amplitude and time scales of the corresponding flares. The variability characteristics of multiple flares have been evaluated for 58 sources of the F-GAMMA sample using the flare decomposition method of \cite{Angelakis2015}. With the addition of an error analysis step, the method consists now of four steps:

\begin{enumerate}
\item {\bf Flare modelling.} This step of the method aims at identifying one basic flare pattern common among
  all events. The operation is executed separately for source and frequency. At first, all the flares in the
  light curve need to be localised by the identification of local maxima. Because flares appear at different
  times and with different amplitudes, the detected events are shifted in time and scaled in flux density so
  that eventually they are all superimposed on the most prominent event. A lower envelope is then fitted to
  the pattern that has resulted from this stacking. It is this envelope that we consider as the template flare
  itself for further analysis.

\item {\bf Correlation.}  This operation aims at finding the optimum time delays between events at different
  frequencies. Instead of using a standard cross-correlation function (e.g. Edelson\& Krolik 1988; Lehar et
  al. 1992), which would treat one pair of light curves each time, we simultaneously include them all. A
  cumulative correlation degree is calculated by multiplying the cross-correlation coefficients of all light
  curve pairs after applying to them different time shifts. The set of time shifts that returns the highest
  degree of cumulative correlation defines the optimum average time delays among frequencies. Clearly, the
  more the available frequencies, the more accurate the estimate of the time shifts is.

\item {\bf Flare Characterisation.} Using the temporal information from the previous step (Correlation), this
  step is meant to identify and characterise the flares that are visible at multiple frequencies, using the
  model from the first step (Flare modelling). The identification of flares at multiple frequencies ensures
  that only significant events are taken into account. Since the frequency availability is not constant the
  number of required frequencies for an event is not strict and it is decided empirically. From the flare
  decomposition, we can calculate variability timescale and amplitude of each flare, which can be used for
  the computation of the variability brightness temperature at each frequency (Eq. 1).

\item {\bf Error analysis.} This operation is meant to provide an estimate of the uncertainty in the flares
  characteristics. Both amplitude and timescales are affected by some degree of uncertainty. This uncertainty
  can be assessed by changing the basic shape of the flare models (both their duration and amplitude) and then
  repeating the flare characterisation using the modified flare models. To each model we associate a goodness
  of fit, provided by the standard deviation of residuals. All models for which this value exceeds by more
  than 10\% the goodness of fit of the best model are disregarded. The range of flare time scales and
  amplitudes for acceptable models set our uncertainty and what we quote as the error of our estimates.
\end{enumerate}

Figures \ref{plt_sim_light} and \ref{plt_sim_light2} shows some examples of simulated light curves (having subtracted the baseline) created after the modeling procedure has been completed. OJ287 (Fig. \ref{plt_sim_light}, upper panel) and 0716+714 (Fig.  \ref{plt_sim_light}, lower panel) are among the fastest sources in our sample; it appears that our method can trace their flux-density variations well. Similarly efficient is the analysis of J0050-0929 (Fig. \ref{plt_sim_light2}, upper panel), which shows slow variability. Less clear is however the case of J0241-0815 (Fig. \ref{plt_sim_light2}, lower panel): although we can trace the variability at individual frequencies well, the significant differences in the variability characteristics at different frequencies makes it hard to efficiently trace the evolution of single flares.

The analyzed sources have been classified according to the quality of their analysis results into three categories: very confident, confident, and less confident. The first category includes sources whose variability characteristics, along with the sampling rate, allow to clearly identify and trace the evolution of flares across all available frequencies. The second category includes sources for which some difficulties have been encountered in modeling the light curves; these difficulties (e.g. a gap in the data, high noise in a minority of frequencies) are expected to have mild effects on the estimation of the variability characteristics. Results for sources of the third category should be regarded as least reliable, because of poor sampling, noisy data, or few available frequencies.

Multi-wavelength modeling of flares provides several advantages over a simple fit. Examining the light curves in different frequencies provides valuable information regarding the evolution of the flares, the type of variability in the source (fast or slow), and the quality of each dataset. This information is taken into account during flare modeling.

In addition, the simultaneous use of all light curves allows us to mitigate the issues related to both the cadence of observations and the superposition of multiple flares. This is obtained by exploiting the general decrease of timescales with frequency. Sources with very fast variability can be best modeled at low frequencies, allowing us to trace back the probable location of flares at high frequency, even below the cadence of observations. On the contrary, flares in slowly varying sources can be best recognized at the highest frequencies; knowing their spectral evolution, we can roughly estimate the contribution of each flare to the variability observed at low frequencies, where, due to the long timescales, single flares cannot easily be isolated. Examples of the multi-wavelength light curves can be found in \cite{Angelakis2010,Angelakis2012,Fuhrmann2016} and in the F-GAMMA website\footnote{http://www3.mpifr-bonn.mpg.de/div/vlbi/fgamma/fgamma.html}.

Given the above considerations, our flare characterization is limited by the cadence of observations at the lowest frequencies. The F-GAMMA sources have a sampling of $\sim 30$ days and in some cases (sources known to show significant variability e.g PKS 0716+714) $\sim 14$ days. Given the typical blazar variability timescales in radio, multi-wavelength information, and the method's ability to mitigate effects of observing cadence, it is rather unlikely any significant event during the F-GAMMA monitoring period has not been accounted for. However, if there are sources in our sample that show variability at timescales significantly shorter than $\sim 14$ days, our results should be treated with caution.

\section{Variability Doppler factors}\label{Variability Doppler factors}
\begin{table*}
\setlength{\tabcolsep}{8pt}
 \centering
 \begin{minipage}{180mm}
 \centering
  \caption{Variability Doppler factors, Lorentz factors and viewing angles for the F-GAMMA sample. Column (1) is the F-GAMMA identification, (2) alternative source name, (3) class (B is for BL Lacs, Q for FSRQs, and G for radio galaxies), (4) redshift, (5) variability Doppler factor, (6) error of the variability Doppler factor (7) Lorentz factor, (8) viewing angle, (9)  mean apparent velocity, (10) number of flares characterized, (11) number of frequencies used for the calculation, (12) frequency that gave the highest estimate of the  variability Doppler factor and (13) confidence on the Doppler factor (0 is for estimates for which we are less confident in our analysis, 1 estimates for which we are confident, and 2 estimates we are very confident in our analysis). }
  \label{tab:variability Doppler factors}
  \begin{tabular}{@{\hskip 0.02cm}ll@{\hskip 0.1cm}ccrcrrrccrr@{}}
  \hline
 F-GAMMA & Alt. name & Class & z & $\delta_{var}$ & $\sigma_{\delta_{var}}$ & $\Gamma$ & $\theta${ (deg.)} & $\beta_{app}$ & No & No & $\nu$ (GHz) & Conf. \\ 
(ID) &  & & & & & &  & & flares & freq. &  & \\
 \hline
J0050$-$0929 & 0048$-$097 & B & 0.634 & 12.8 & 3.4 & - & - & - & 7 & 9 & 2.64 & 2\\ 
J0102+5824 & 0059+5808 & Q & 0.644 & 21.9 & 3.6 & 12.0 & 1.5 & 6.89 & 8 & 9 & 2.64 & 2  \\ 
J0136+4751 & 0133+476 & Q & 0.859 & 13.7 & 2.7 & 9.5 & 3.7 & 8.43 & 4 & 7 & 8.35 & 2 \\ 
J0217+0144 & PKS0215+015 & Q & 1.715 & 27.1 & 1.3 & 19.1 & 1.9 & 17.30 & 7 & 9 & 2.64 & 2 \\ 
J0222+4302 & B0219+428 & B & 0.444 & 4.3 & 0.2 & - & - & - & 3 & 9 & 2.64 & 1 \\ 
J0237+2848 & 0234+285 & Q & 1.206 & 12.2 & 4.3 & 14.9 & 4.6 & 14.65 & 4 & 9 & 2.64 &2  \\ 
J0238+1636 & 0235+164 & B & 0.940 & 29.0 & 7.7 & 14.6 & 0.3 & 2.00 & 7 & 9 & 2.64 &2 \\ 
J0241$-$0815 & 0238$-$084 & G & 0.005 & 0.3 & 0.0 & 2.0 & 26.9 & 0.22 & 6 & 8 & 8.35 &0 \\ 
J0336+3218 & PKS0333+321 & Q & 1.259 & 2.9 & 0.2 & 21.4 & 10.0 & 10.66 & 4 & 2 & 86.00 &0  \\ 
J0339$-$0146 & 0336$-$019 & Q & 0.850 & 16.7 & 3.5 & 12.2 & 3.2 & 11.34 & 5 & 9 & 4.85 & 0 \\ 
J0359+5057 & 0355+50 & Q & 1.520 & 26.3 & 3.6 & 13.2 & 0.2 & 1.39 & 4 & 10 & 2.64 & 1 \\  
J0418+3801 & B20415+37 & G & 0.049 & 2.0 & 0.4 & 7.0 & 20.0 & 4.85 & 6 & 9 & 2.64 & 2 \\ 
J0423$-$0120 & 0420$-$014 & Q & 0.916 & 43.9 & 9.2 & 22.2 & 0.3 & 4.44 & 8 & 9 & 2.64 & 1 \\ 
J0433+0521 & 0430+052 & G & 0.033 & 2.1 & 0.1 & 6.8 & 19.9 & 4.81 & 4 & 9 & 4.85 & 1 \\ 
J0530+1331 & 0528+134 & Q & 2.070 & 12.9 & 2.5 & 10.8 & 4.4 & 10.50 & 5 & 7 & 8.35 & 2 \\ 
J0654+4514 & S40650+453 & Q & 0.928 & 13.8 & 2.6 & - & - & - & 6 & 9 & 14.6 & 2 \\ 
J0719+3307 & TXS0716+332 & Q & 0.779 & 14.1 & 0.5 & - & - & - & 5 & 7 & 2.64 & 0 \\ 
J0721+7120 & 0716+714 & B & 0.328 & 14.0 & 0.9 & 10.8 & 3.9 & 10.22 & 14 & 10 & 2.64 &2  \\ 
J0730$-$1141 & PKS0727$-$115 & Q & 1.591 & 39.8 & 6.9 & - & - & - & 7 & 9 & 2.64 & 2 \\ 
J0738+1742 & 0735+178 & B & 0.424 & 4.5 & 0.4 & 3.6 & 12.2 & 3.30 & 3 & 9 & 4.85 & 1  \\ 
J0808$-$0751 & 0805$-$077 & Q & 1.837 & 14.9 & 1.2 & 24.3 & 3.5 & 22.42 & 4 & 9 & 8.35 & 2 \\ 
J0818+4222 & 0814+425 & B & 0.530 & 7.8 & 2.6 & 4.1 & 3.2 & 1.72 & 5 & 9 & 23.05  & 1\\ 
J0824+5552 & S40820+560 & Q & 1.417 & 2.4 & 0.5 & - & - & - & 4 & 2 & 86.00   & 0\\ 
J0841+7053 & 0836+710 & Q & 2.218 & 12.1 & 0.0 & 19.0 & 4.4 & 17.69 & 1 & 9 & 2.64 & 0  \\ 
J0854+2006 & 0851+202 & B & 0.306 & 8.7 & 1.1 & 7.6 & 6.6 & 7.49 & 10 & 10 & 4.85 & 2 \\ 
J0920+4441 & S40917+449 & Q & 2.190 & 5.0 & 0.9 & 2.8 & 6.2 & 1.45 & 1 & 8 & 4.85 & 0 \\ 
J0958+6533 & 0954+658 & B & 0.367 & 10.7 & 1.7 & 7.9 & 5.0 & 7.31 & 9 & 9 & 2.64 & 2\\ 
J1104+3812 & PKS1101+384 & B & 0.030 & 1.7 & 0.1 & 1.1 & 8.6 & 0.14 & 6 & 9 & 2.64 & 2 \\ 
J1130$-$1449 & 1127$-$145 & Q & 1.184 & 21.9 & 0.0 & 13.0 & 1.9 & 9.46 & 4 & 8 & 4.85  & 1\\ 
J1159+2914 & PKS1156+295 & Q & 0.725 & 12.8 & 0.0 & 16.6 & 4.3 & 16.13 & 6 & 9 & 2.64 & 2 \\ 
J1217+3007 & PKS1215+303 & B & 0.130 & 1.1 & 0.3 & 1.0 & 20.7 & 0.03 & 3 & 8 & 86.00  & 0\\ 
J1221+2813 & QSOB1219+285 & B & 0.102 & 2.6 & 0.6 & 4.6 & 19.8 & 4.08 & 5 & 8 & 2.64  & 1\\ 
J1229+0203 & 1226+023 & Q & 0.158 & 3.7 & 1.0 & 12.0 & 11.3 & 8.58 & 6 & 8 & 8.35  & 1\\ 
J1256$-$0547 & 1253$-$055 & Q & 0.536 & 16.8 & 2.9 & 12.3 & 3.2 & 11.42 & 9 & 9 & 23.05 & 1 \\ 
J1310+3220 & 1308+326 & B & 0.997 & 15.8 & 1.7 & 17.1 & 3.6 & 16.99 & 6 & 9 & 2.64  & 1\\ 
J1332$-$0509 & PKS1329$-$049 & Q & 2.150 & 18.9 & 3.9 & 11.1 & 2.1 & 7.70 & 6 & 8 & 4.85  & 1\\ 
J1504+1029 & 1502+106 & Q & 1.839 & 17.3 & 2.7 & 11.4 & 2.8 & 9.63 & 5 & 10 & 4.85  & 2\\ 
J1512$-$0905 & 1510$-$089 & Q & 0.360 & 12.3 & 2.8 & 19.0 & 4.4 & 17.76 & 10 & 9 & 2.64  & 2\\ 
J1613+3412 & 1611+343 & Q & 1.400 & 2.4 & 0.5 & 31.5 & 9.2 & 11.93 & 3 & 2 & 86.00  & 0\\ 
J1635+3808 & 1633+382 & Q & 1.814 & 20.3 & 2.8 & 14.9 & 2.6 & 13.78 & 8 & 10 & 8.35  & 2\\ 
J1642+3948 & 1641+399 & Q & 0.593 & 10.4 & 2.9 & 11.3 & 5.5 & 11.22 & 6 & 9 & 2.64   &2\\ 
J1653+3945 & 1652+398 & B & 0.033 & 2.1 & 0.0 & 1.3 & 8.2 & 0.24 & 5 & 9 & 2.64  & 0\\ 
J1733$-$1304 & PKS1730$-$130 & Q & 0.902 & 17.6 & 3.4 & 15.3 & 3.2 & 15.09 & 7 & 10 & 42.00 & 0  \\ 
J1751+0939 & 1749+096 & B & 0.322 & 14.2 & 2.0 & 7.8 & 2.3 & 4.36 & 9 & 9 & 2.64  & 2\\ 
J1800+7828 & 1803+784 & B & 0.680 & 21.2 & 5.0 & 10.8 & 0.6 & 2.53 & 8 & 9 & 2.64 & 2 \\ 
J1824+5651 & 1823+568 & B & 0.664 & 1.0 & 0.2 & 34.8 & 13.4 & 8.36 & 5 & 1 & 86.00  & 0\\ 
J1848+3219 & TXS1846+322 & Q & 0.798 & 12.1 & 1.4 & 7.0 & 3.2 & 4.69 & 7 & 9 & 2.64  & 2\\ 
J1849+6705 & S41849+670 & Q & 0.657 & 8.1 & 1.4 & 17.0 & 6.0 & 14.48 & 6 & 9 & 4.85 & 1 \\ 
J2025$-$0735 & PKS2022$-$077 & Q & 1.388 & 16.5 & 4.9 & 24.6 & 3.3 & 23.20 & 7 & 9 & 2.64  & 2\\ 
J2143+1743 & PKS2141+175 & Q & 0.213 & 8.8 & 1.8 & 4.7 & 3.0 & 2.15 & 8 & 9 & 2.64  & 2\\ 
J2147+0929 & 2144+092 & Q & 1.113 & 13.6 & 1.8 & - & - & - & 5 & 9 & 2.64  & 2\\ 
J2202+4216 & 2200+420 & B & 0.069 & 6.1 & 0.8 & 5.6 & 9.4 & 5.49 & 12 & 9 & 2.64 & 2  \\ 
J2203+1725 & PKS2201+171 & Q & 1.076 & 10.0 & 0.5 & 8.4 & 5.7 & 8.17 & 4 & 9 & 8.35 & 0 \\ 
J2203+3145 & 2201+315 & Q & 0.295 & 4.1 & 1.1 & 6.3 & 13.2 & 5.79 & 8 & 5 & 23.05  & 0\\ 
J2229$-$0832 & 2227$-$088 & Q & 1.560 & 21.0 & 0.6 & 10.6 & 0.5 & 1.92 & 5 & 9 & 4.85  & 1\\ 
J2232+1143 & 2230+114 & Q & 1.037 & 15.1 & 4.8 & 8.1 & 1.9 & 4.00 & 8 & 9 & 4.85  & 1\\ 
J2253+1608 & 2251+158 & Q & 0.859 & 17.0 & 3.7 & 10.4 & 2.6 & 7.90 & 7 & 10 & 42.00  & 2\\ 
J2327+0940 & PKS2325+093 & Q & 1.841 & 17.2 & 2.3 & - & - & - & 6 & 9 & 4.85  & 2\\ 
\hline
\end{tabular}
\end{minipage}
\end{table*}

\begin{figure}
\resizebox{\hsize}{!}{\includegraphics[scale=1]{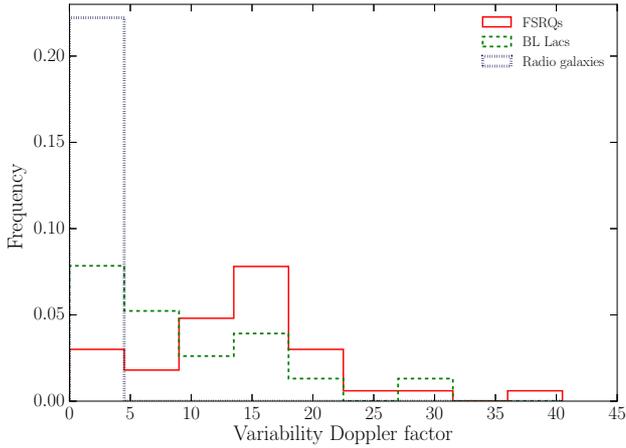} }
 \caption{Distribution of the variability Doppler factor for the F-GAMMA sources. Solid red is for the FSRQs, dashed green for the BL Lacs, and dotted blue for the radio galaxies in our sample.}
 \label{plt_hist_dvar}
 \end{figure}
\begin{figure}
\resizebox{\hsize}{!}{\includegraphics[scale=1]{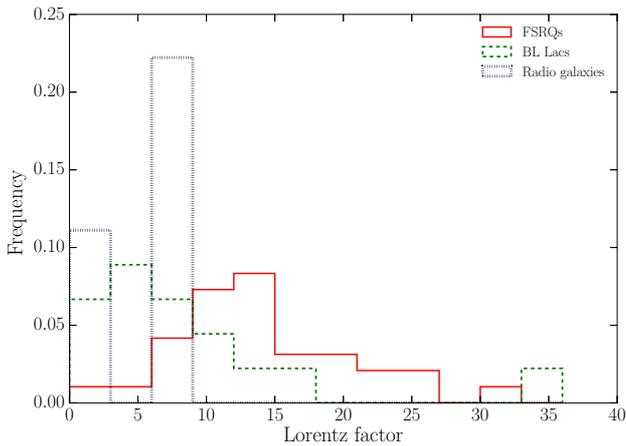} }
 \caption{Distribution of the Lorentz factor for the F-GAMMA sources. Solid red is for the FSRQs, dashed green for the BL Lacs, and dotted blue for the radio galaxies in our sample.}
 \label{plt_hist_lorentz}
 \end{figure}
\begin{figure}
\resizebox{\hsize}{!}{\includegraphics[scale=1]{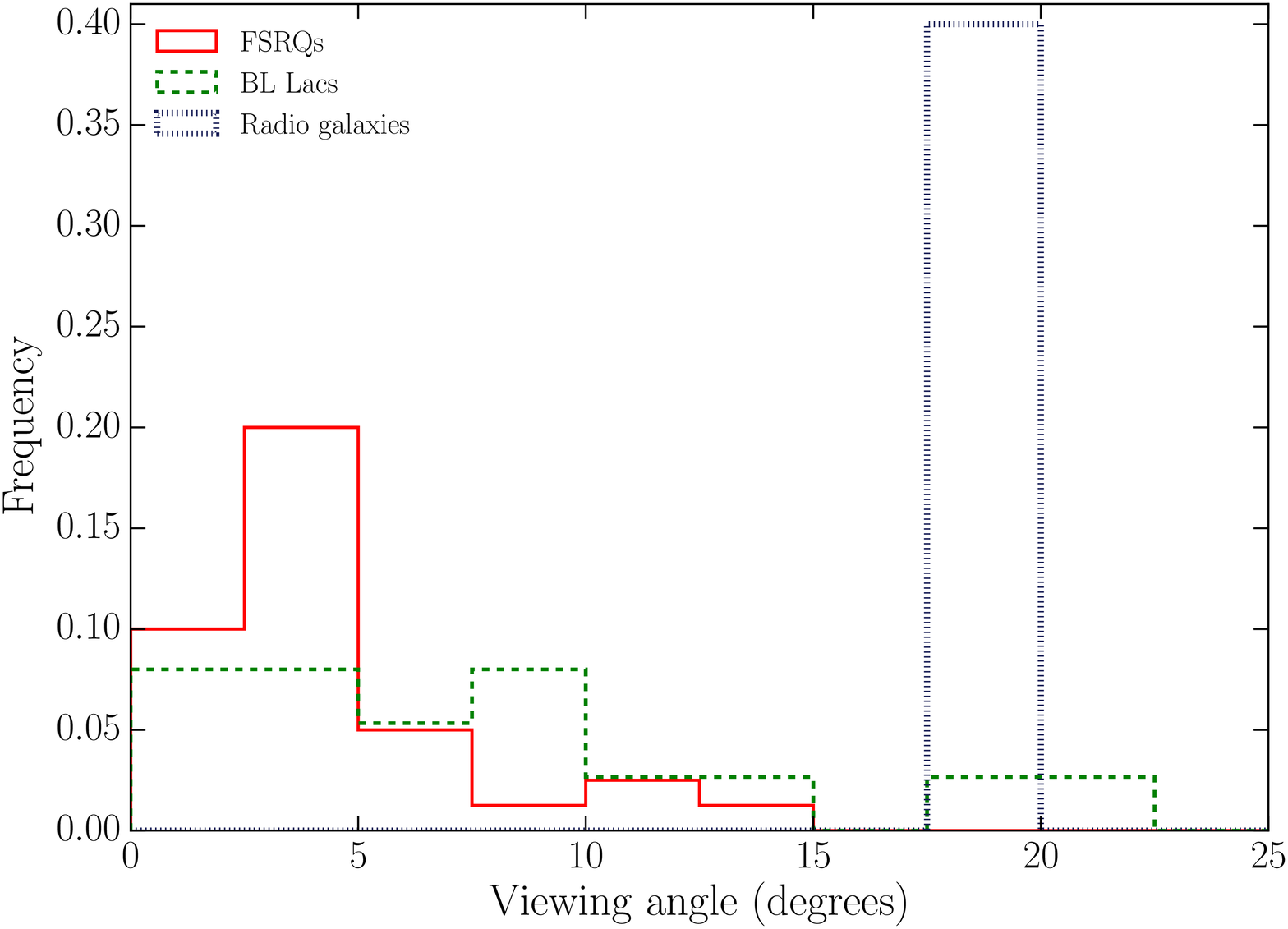} }
 \caption{Distribution of the viewing angle for the F-GAMMA sources. Solid red is for the FSRQs, dashed green for the BL Lacs, and dotted blue for the radio galaxies in our sample.}
 \label{plt_hist_viewing_angle}
 \end{figure}
\begin{figure}
\resizebox{\hsize}{!}{\includegraphics[scale=1]{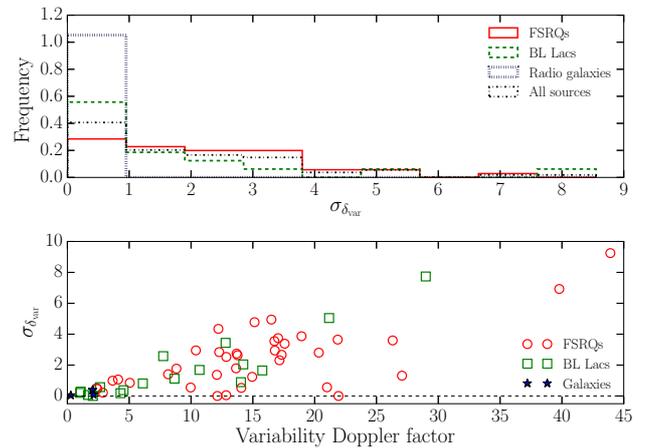}  }
 \caption{Upper panel: Distribution of the error estimates of the variability Doppler factor for the F-GAMMA sources . Solid red is for the FSRQs, dashed green for the BL Lacs, and dotted blue for the radio galaxies, while black dash-dot for the whole sample. Lower panel: Variability Doppler factor versus the error of each estimate. Red circle is for the FSRQs, green square for the BL Lacs, and blue star for the radio galaxies.}
 \label{plt_multi-plt_error}
 \end{figure}

Once the flares have been identified and modeled, their variability characteristics can be estimated. Their amplitude coincides with the flux density at the peak. The timescales of a flare are the time spans between the beginning of the flare and its peak, and between its peak and its end. We define the beginning of a flare as the time at which its flux density exceeds a threshold of 0.25 times the average uncertainty in the flux density measurements. The end of a flare is defined similarly, as the time when the flux density drops below that threshold. This definition helps dealing with flares that extend to very long timescales without carrying any significant contribution to the total flux density.

Through the variability characteristics of flares, the associated variability brightness temperatures can be calculated using the following formula (Eq. \ref{tvar_num}).
\begin{equation}
T_\mathrm{var}=1.47 \cdot 10^{13}\frac{D^2_L \Delta S_\mathrm{ob}(\nu)}{\nu^2t^2_\mathrm{var}(1+z)^4},
\label{tvar_num}
\end{equation}
where $T_\mathrm{var}$ is the variability brightness temperature in Kelvin, $D_L$ is the luminosity distance in Mpc, $S_\mathrm{ob}(\nu)$ the flux density in Jy, $t_\mathrm{var}$ the variability timescale in days, $z$ is the redshift, and $\nu$ the observing frequency in GHz. The numerical factor is related to units and the geometry of the emitting region. Assuming that while flaring, sources reach equipartition \citep{Readhead1994}, the intrinsic brightness temperature will be equal to the equipartition brightness temperature $T_\mathrm{eq}=5\times 10^{10} K$  \citep{Readhead1994,Lahteenmaki1999-II}. By comparing the observed and intrinsic brightness temperature we estimate the variability Doppler factor as follows:
\begin{equation}
\delta_\mathrm{var}=(1+z)\sqrt[3]{\frac{T_\mathrm{var}}{T_\mathrm{eq}}}.
\label{multi-varia-Doppler}
\end{equation}
For the full derivation of Eq. \ref{multi-varia-Doppler} see appendix \ref{app:Doppler_derivation}. The highest variability brightness temperature observed in a source provides the highest constrain to the variability Doppler factor. The highest estimate for the variability Doppler factor found in each source is the estimate we quote in Table \ref{tab:variability Doppler factors}.

We calculate the Lorentz factor ($\Gamma_{var}$) and viewing angle ($\theta_{var}$) using Eq. \ref{lorentz}, \ref{theta} and the apparent velocity ($\beta_\mathrm{app}$). In order to estimate the mean $\beta_\mathrm{app}$, we use data from the MOJAVE survey \citep{Lister2005}, for all our sources with available estimates in the literature \citep{Lister2009-2,Lister2013}.
\begin{equation}
\Gamma_\mathrm{var} = \frac{\beta_\mathrm{app}^2 + \delta_\mathrm{var}^2 + 1}{2\delta_\mathrm{var}},
\label{lorentz}
\end{equation}
\begin{equation}
\theta_\mathrm{var} = \arctan \frac{2\beta_\mathrm{app}}{\beta_\mathrm{app}^2 + \delta_\mathrm{var}^2-1}.
\label{theta}
\end{equation}

All the estimates for the Doppler factors as well as Lorentz factors and viewing angles are summarized in Table \ref{tab:variability Doppler factors}. It is obvious that the more flares and frequencies used for the characterization of the light curves, the better we can constrain the variability brightness temperature, and the more confident we are about the results of our analysis. The number on the last column of Table \ref{tab:variability Doppler factors} (column 13) denotes our confidence on the estimate. $0$ is for the cases that we are less confident in the results of our analysis, $1$ is for the cases we are confident, and $2$ is for the cases we are very confident. The confidence in the estimate of the Doppler factor depends on the abundance of data points available for each source. Sparse data, large observational gaps or fewer available frequencies could severely hamper tracking the evolution and characterization of the flares which is the basis of our methodology. Such problems in the analysis could lead to the underestimation of the variability brightness temperature. An additional cause for our lack in confidence would be a general lack of flares in a source. For a discussion and notes on the analysis of individual sources see {\color{blue}Marchili et al. (in prep.)}. A more conservative approach to the equipartition brightness temperature would be to use the inverse Compton catastrophe limit $T_{IC}=10^{12} K$  \citep{Kellerman1969}. This would bring our estimates lower by a factor of $\sim 2.7$. We chose to use the equipartition limit since the variability Doppler factors using equipartition \citep{Hovatta2009} best describe the blazar populations \citep{Liodakis2015-II}.

Figure \ref{plt_hist_dvar} shows the distribution of the variability Doppler factors,  Fig. \ref{plt_hist_lorentz} the distribution of Lorentz factors, and Fig. \ref{plt_hist_viewing_angle} the distribution of viewing angles for the F-GAMMA sources, where solid red is for the FSRQs, dashed green for the BL Lacs and dotted blue for the radio galaxies.  FSRQs and BL Lacs (except two BL Lacs) have viewing angles lower than 15 degrees, consistent with the current view of blazars \citep{Ghisellini1993,Urry1995}. The values for the mean and std of the populations are summarized in Table \ref{tab:pop_char}.
\begin{table}
\setlength{\tabcolsep}{11pt}
\centering
  \caption{Mean and standard deviation (std) of the Doppler factors, Lorentz factors and viewing angles for the three populations in our sample.}
  \label{tab:pop_char}
\begin{tabular}{@{}cccc@{}}
 \hline
      & FSRQ & BL Lacs & Radio galaxies \\
  \hline
   $\delta_{var}$  &    &  \\
  mean  &  15.21  & 9.2 & 1.4  \\
  std  &  8.7 & 7.6 & 0.8 \\
  \hline
  $\Gamma_{var}$ &    & & \\
mean  &  13.9  & 9.2 & 5.2 \\
  std  &  6.1 & 8.8 &  2.2 \\
  \hline
   $\theta_{var}$ (degrees) &    & & \\
  mean  &  4.2  & 7.8 & 22.2 \\
  std  &  2.9  & 6.1 & 3.3 \\ 
\hline
\end{tabular}
\end{table}

The FSRQs appear to have higher Doppler factors than the BL Lacs, and both higher than the radio galaxies as expected. The same is the case for the Lorentz factors and the opposite for the viewing angles. The highest Lorentz factor is attributed to the BL Lac object J1824+5651; this estimate, however, falls within the category of the ``less confident'' results. The high Lorentz factor that we found for this source may be caused by an underestimation of the Doppler factor ($\delta_{var}=1$). A higher Doppler factor estimate (even $\delta_{var}=2$) would bring the value of the Lorentz factor lower than that of the fastest FSRQs with estimates labeled ``confident'' and ``very confident''  (J0423$-$0120 $\Gamma_{var}=22.2$ and J2025$-$0735 $\Gamma_{var}=24.6$ respectively).

 The mean value for the Doppler factor for the ``confident'' and ``very confident'' estimates ($\sim 14$) is very similar to the overall mean ($\sim 12$). Thus we can conclude that the reliability of the estimates (in these two categories) does not strongly influence the results of our analysis However for the ``less confident'' estimates, the mean is $\sim 6$ while the mean of the apparent velocity (7.6) is similar to the sample mean (8.3) as is the case for the ``confident'' and ``very confident'' categories. The resulting $\Gamma_{var}$ and $\theta_{var}$ for the ``less confident'' estimates are larger than that of the sample. There are two possible explanations for this discrepancy. Either there are indeed unaccounted for peculiarities of the analysis which lead to underestimating the Doppler factor for these sources or the majority of the sources labeled ``less confident'' are slowly variable. In the latter case, their Doppler factors will be low, causing an increase of the value of the Lorentz factor. In that case, the source composition of the category is biasing the results. In either case, estimates labeled as ``less confident'' should be treated with caution.

In order to asses the significance in possible differences between FSRQs and BL Lacs in our sample, we use the Wilcoxon rank-sum test which gives the probability of two samples to have been drawn from the same distribution (the alternative hypothesis is that values from one sample are more likely to be larger than the other). The probability of the two samples being drawn from the same distribution is 1.1\% for the variability Doppler factors, 0.3\% for the Lorentz factors, and 4.9\% for the viewing angles. Although we cannot reach solid conclusions for the populations, this would imply that FSRQs have on average higher Doppler factors, Lorentz factors, and smaller viewing angles than the BL Lacs \citep{Jorstad2005,Hovatta2009,Lister2013,Liodakis2015}.

Figure \ref{plt_multi-plt_error} shows the distribution of the errors in our estimates (upper panel) and the error of each estimate against the value of the Doppler factor (lower panel). The mean of the error distribution for the whole sample is 2.07 with a standard deviation of 1.99. The highest percentage error in our estimates is 35.5\%, which is comparable to the most accurate estimates available in the literature ($\sim$30\% error on average) as derived from population models. Overall our method has a 16\% error on average, making  our method the most accurate approach to date.

\section{Comparison with other methods}\label{compar}
\begin{figure}
\resizebox{\hsize}{!}{\includegraphics[scale=1]{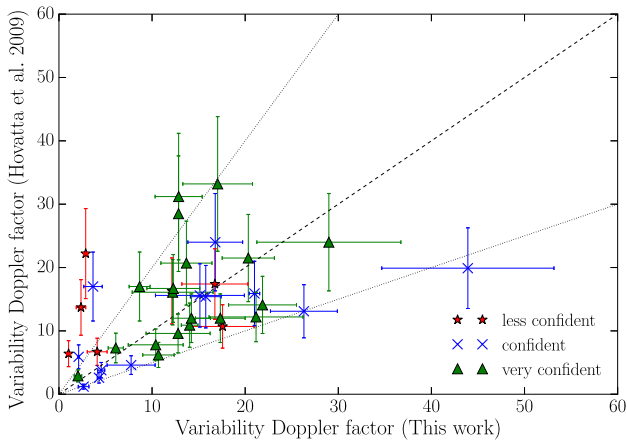}  }
 \caption{Variability Doppler factors (this work) versus the variability Doppler factors from \citet{Hovatta2009}.  The green triangle is for sources for which we are very confident of our analysis, blue x for the sources we are confident, and red star for the sources we are less confident (see Table \ref{tab:variability Doppler factors}). The dashed line denotes the $y=x$ line, whereas the dotted lines mark the factor of two envelope. The error of the $y$-axis is the 30\% average error derived through population modeling.}
 \label{plt_dvar_vs_Hov}
 \end{figure}

The F-GAMMA sample is not flux-limited or complete, and hence our results (drawn form it) cannot be statistically tested against blazar population models. We can, however, use as a proxy estimates that have been shown to be consistent with the population. We chose to compare our Doppler factors with \cite{Hovatta2009} for two reasons: (a) it is the most recent study on the variability Doppler factors using a different approach for estimating the variability brightness temperature; (b) estimates from \cite{Hovatta2009} have been tested against population models \citep{Liodakis2015-II} and it was shown that they can adequately describe both the FSRQ and BL Lac populations. 

Figure \ref{plt_dvar_vs_Hov} shows the comparison between the variability Doppler factors derived in this work, and the variability Doppler factors from \cite{Hovatta2009}. The two samples have 38 sources in common. In \cite{Hovatta2009} the authors comment on how difficult is to determine the exact error of the $\delta_{var}$. They provide an upper limit to the error of their estimates by calculating the standard deviation of the different $\delta_{var}$ for individual well-defined flares in each source. They find a median standard deviation of  $\sim$27\%. However, populations models find that the on-average error of their estimates is $\sim$30\%. The error of the $y$-axis is the $\sim$30\% average error derived from population modeling \citep{Liodakis2015-II}. 

Although there are some discrepancies, the majority of the estimates are within the factor of two envelope and most are within errors. The two-sample Kolmogorov-Smirnov test (K-S test) yielded a 69.25\% probability of consistency between the estimates of the two methods (the null hypothesis is that the two samples are drawn from the same population). Testing for their correlation, the Spearman rank-order correlation yielded a correlation coefficient $r$=0.5 (-1 negative correlation, 0 no correlation, 1 positive correlation) with a $\sim 0.1\%$ probability of the two samples being uncorrelated. Excluding the estimates for which we are less confident in our analysis, the K-S test yielded a 94.4\% probability of consistency, and the Spearman rank-order a $r$=0.57 with a $\sim 0.08\%$ probability of uncorrelated samples. Thus we can conclude that the estimates of the two methods are drawn from the same population.

Our method tends to yield higher estimates of the Doppler factors than those of \cite{Hovatta2009}, although this is not confirmed by the Wilcoxon rank-sum test (41.18\%). This trend is more prominent at high values of the Doppler factor, which is to be expected since in our approach the effects of cadence of observations are mitigated. On the other hand there are sources for which our estimates are lower. Although some of these estimates fall under our ``less confident'' category, there are sources for which we are confident in our results. A possible explanation of this discrepancy would be the uncertainty in the estimates in \cite{Hovatta2009}. For example, an inadequate fit or fitting what would appear as a flaring event but is instead stochastic variability, could lead to underestimating the timescale and consequently overestimating the Doppler factor. A more probable scenario, given the span (roughly $\sim$35 years) of the Mets\"ahovi monitoring program, would be the occurrence of a major flare in each of these sources outside the F-GAMMA monitoring period. This could lead to higher brightness temperatures and hence higher Doppler factors for these sources. The origin of this discrepancy needs to be investigated on the basis of a source-by-source analysis, which is currently in progress. In any case, inconsistencies only concern 9 sources. Their impact on the results of this study is therefore very low.

\section{Summary}\label{summ}

We used specially designed algorithms in order to identify, track and characterize flares throughout a large number of radio frequencies from 2.64 up to 142.33 GHz with data from the F-GAMMA blazar monitoring program \citep{Fuhrmann2007,Angelakis2010,Angelakis2012,Fuhrmann2016}.  Using the variability brightness temperature obtained with this approach \citep{Angelakis2015} we were able to calculate the variability Doppler factor (Eq. \ref{multi-varia-Doppler}) for 58 sources, for 20 of which no variability Doppler factor had been estimated before, and provide error estimates on a source-by-source basis. Combined with apparent velocities from the MOJAVE survey \citep{Lister2005} we calculated the Lorentz factor and viewing angles for 50 sources. All values, as well as additional information on the sources, are listed in Table \ref{tab:variability Doppler factors}. Our results can be summarized as follows.

\begin{itemize}

\item There are differences in the Doppler factor estimates between the BL Lacs and FSRQs. FSRQs appear to have significantly larger Doppler factors and Lorentz factors and smaller viewing angles consistent with our current understanding of blazars. \citep{Jorstad2005,Hovatta2009,Lister2013,Liodakis2015}.

\item Both FSRQ and BL Lac populations have higher Doppler and Lorentz factors than the radio galaxies. The viewing angles are typically $<15$ degrees for all blazars but one BL Lac object, whereas radio galaxies have viewing angles $\geq 20$ degrees consistent with our current view on the unification of radio galaxies \citep{Ghisellini1993,Urry1995}.

\item The mean error of our estimates is 2.07. Our highest percentage error (35.5\%) is comparable to the most accurate estimates available in the literature (30\% on average, \citealp{Hovatta2009}), whereas our on average error is 16\%. Thus, our method is the most accurate for estimating the Doppler factor of blazar jets to date, with the unique ability to provide error estimates on a source-by-source basis.

\item We compared the Doppler factors derived from this work to estimates from the literature \citep{Hovatta2009} that have been shown to adequately describe the blazar populations \citep{Liodakis2015-II}. There are very few discrepancies which can be attributed either to uncertainties in the analysis of the literature values or in the analysis presented here. Nevertheless, the two samples are consistent within the errors, as is validated confidently by the Kolmogorov-Smirnov and the Spearman rank-order correlation tests.
\end{itemize}

The multi-wavelength variability Doppler factors presented here were found to be consistent with the estimates in \citep{Hovatta2009} that can adequately describe the FSRQ and BL Lac populations \citep{Liodakis2015-II}. Hence, we can conclude that they are not only the most accurate estimates yet, but can also describe blazars as a population, validating our results and stressing the importance and wealth of information that can be obtained from multi-wavelength monitoring programs such as the F-GAMMA.

\section*{Acknowledgments}
The authors would like to thank Talvikki Hovatta, Shoko Koyama, and the anonymous referee for comments and suggestions that helped improve this work. This research was supported by the ``Aristeia'' Action of the  ``Operational Program Education and Lifelong Learning'' and is co-funded by the European Social Fund (ESF) and Greek National Resources, and by the European Commission Seventh Framework Program (FP7) through grants PCIG10-GA-2011-304001 ``JetPop'' and PIRSES-GA-2012-31578 ``EuroCal''. This research has made use of data from the MOJAVE database that is maintained by the MOJAVE team \citep{Lister2009-2}.  Our study is based on observations carried out with the 100 m telescope of the MPIfR (Max-Planck-Institut f\"{u}r Radioastronomie) and the IRAM 30 m telescope. IRAM is supported by INSU/CNRS (France), MPG (Germany) and IGN (Spain). I. N., I.M. and V.K. were supported for this research through a stipend from the International Max Planck Research School (IMPRS) for Astronomy and Astrophysics at the Universities of Bonn and Cologne.

\appendix

\section{Variability Doppler factor Derivation}\label{app:Doppler_derivation}

We use the expression for the variability brightness temperature (Eq. \ref{tvar}) from \cite{Blandford1979}, and Eq. \ref{intens} and \ref{intens_over_nu}  in order to obtain the correct expression for the variability Doppler factor.
\begin{equation}
T_\mathrm{var}=\frac{D^2_\mathrm{L} \Delta S_\mathrm{ob}(\nu)}{2\nu^2t^2_\mathrm{var}k(1+z)^4},
\label{tvar}
\end{equation}
\begin{equation}
I(\nu)=\frac{2k\nu^2T_\mathrm{var}}{c^2}=\frac{\Delta S(\nu)}{\theta^2},
\label{intens}
\end{equation}
\begin{equation}
\frac{I'(\nu)}{\nu'^3}=\frac{I(\nu)}{\nu^3},
\label{intens_over_nu}
\end{equation}
where $T_\mathrm{var}$ is the variability brightness temperature, $D_L$ the luminosity distance, $\nu$ is the frequency, $t_\mathrm{var}$ the variability timescale, z the redshift, $I(\nu)$ the intensity, $k$ Boltzman's constant, $c$ the speed of light, $ \Delta S(\nu)$ the flux density, and $\theta$ the angular size of the source. Primed symbols denote rest-frame quantities. Combining equations \ref{tvar} and \ref{intens}:
\begin{equation}
T_\mathrm{var}=\frac{D^2_L I(\nu)\theta^2}{2\nu^2t^2_\mathrm{var}k(1+z)^4}.
\label{tvar2}
\end{equation}
The observed transverse size is:
\begin{equation}
R=\frac{\delta_\mathrm{var} ct_\mathrm{var}}{1+z}=D_A\theta\Rightarrow\theta=\frac{\delta_\mathrm{var} ct_\mathrm{var}}{D_A(1+z)},
\end{equation}
where $D_\mathrm{A}$ is the angular diameter distance to the source and $\delta_\mathrm{var}$ the Doppler factor. From Eq. \ref{intens_over_nu} we have that $I(\nu)=\delta_\mathrm{var}^3I'(\nu)$. If we take cosmological expansion into account $I(\nu)=\delta_\mathrm{var}^3I'(\nu)(1+z)^{-3}$ (because $\nu'=(1+z)\nu)$). Putting everything in equation \ref{tvar2} then:

\begin{equation}
T_\mathrm{var}=\frac{D^2_L I(\nu)}{2\nu^2t^2_\mathrm{var}k(1+z)^4}\left(\frac{\delta_\mathrm{var} ct_{var}}{D_A(1+z)}\right)^2.
\end{equation}
The angular diameter distance is defined as $D_A=\frac{D}{1+z}$ and the luminosity distance as $D_L=D(1+z)\Rightarrow D^2=\frac{D_L^2}{(1+z)^2}$. The variability brightness temperature becomes,
\begin{eqnarray}
T_\mathrm{var}&=&\frac{c^2}{2k}\frac{D^2I(\nu)}{\nu^2t^2_\mathrm{var}(1+z)^2}\frac{\delta_\mathrm{var}^2 c^2t_\mathrm{var}^2}{D^2}\nonumber\\
&=&\frac{c^2}{2k}\frac{I(\nu)}{\nu^2(1+z)^2}\delta_\mathrm{var}^2 c^2\nonumber\\
&=&\frac{c^2}{2k}\frac{I'(\nu)\delta_\mathrm{var}^3\delta_\mathrm{var}^2}{\nu'^2\delta_\mathrm{var}^2(1+z)^3(1+z)^2(1+z)^{-2}}\nonumber\\
&=&\frac{c^2I'(\nu)}{2kv'^2}\frac{\delta_\mathrm{var}^3}{(1+z)^3}=\frac{\delta_\mathrm{var}^3}{(1+z)^3}T'_\mathrm{var}
\end{eqnarray}
Assuming that while flaring the source reaches equipartition between the energy density of the magnetic field and that of the radiating particles \citep{Readhead1994} we can substitute the intrinsic brightness temperature with the equipartition brightness temperature ($T_{eq}=5\times 10^{10} K$ ),
\begin{equation}
T_{var}=\frac{\delta_{var}^3}{(1+z)^3}T_{eq}
\end{equation}
The variability Doppler factor will be:
\begin{equation}
\delta_{var}=(1+z)\sqrt[3]{\frac{T_{var}}{T_{eq}}}
\end{equation}

\bibliographystyle{mnras}
\bibliography{bibliography} 

\label{lastpage}

\end{document}